\documentclass[12pt]{article}
\usepackage{amsfonts,latexsym,amscd,amsmath,theorem,amssymb}
\usepackage{epsfig}
\input diagrams.sty

\newcommand{\R}{{\mathbb{R}}}
\newcommand{\Z}{{\mathbb{Z}}}

\newcommand{\C}{{\mathbb{C}}}
\newcommand{\I}{{\mathbb{I}}}

\newcommand{\CP}{{\mathbb{C}}{{P}}}

\newcommand{\beq}{\begin{equation}}
\newcommand{\eeq}{\end{equation}}
\newcommand{\bea}{\begin{eqnarray}}
\newcommand{\eea}{\end{eqnarray}}
\newcommand{\ra}{\rightarrow}

\newcommand{\cd}{\partial}
\newcommand{\wt}{\widetilde}

\newcommand{\ms}{{\sf M}}

\newcommand{\g}{{\mathfrak{g}}}
\newcommand{\kk}{{\mathfrak{k}}}
\newcommand{\p}{{\mathfrak{p}}}
\newcommand{\Q}{{\mathcal{Q}}}

\newcommand{\Leff}{L_{\rm eff}}

\newcommand{\nvec}{{\mbox{\boldmath{${\rm n}$}}}} 
\newcommand{\xvec}{{\mbox{\boldmath{${\rm x}$}}}} 
\newcommand{\evec}{\mbox{\boldmath{${\rm e}$}}} 

\newcommand{\tr}{{\rm tr}\, }

\newcommand{\diag}{{\rm diag}}

\newcommand{\free}[2]{{\rm Free}(#1,#2)}

\theoremstyle{plain}
\newtheorem{thm}{Theorem}

\newtheorem{cor}[thm]{Corollary}

{\theorembodyfont{\rmfamily}

}
\newcommand{\news}{\setcounter{equation}{0}}

\begin{document}

\title{Fermionic quantization of Hopf solitons}
\author{
S. Krusch\thanks{E-mail: {\tt S.Krusch@kent.ac.uk}}\\
Institute of Mathematics, University of Kent\\
Canterbury CT2 7NF, England\\ \\
J.M. Speight\thanks{E-mail: {\tt speight@maths.leeds.ac.uk}}\\
Department of Pure Mathematics, University of Leeds\\
Leeds LS2 9JT, England
}

\date{}
\maketitle

\begin{abstract}
In this paper we show how to quantize Hopf solitons using the
Finkelstein-Rubinstein approach. Hopf solitons can be
quantized as fermions if their Hopf charge is odd. Symmetries of classical
minimal energy configurations induce loops in configuration space which
give rise to constraints on the wave function. These constraints depend on
whether the given loop is contractible. 
Our method is to exploit the relationship between the configuration 
spaces of the Faddeev-Hopf and Skyrme models provided by the Hopf fibration.
We then use recent results in the
Skyrme model to determine whether loops are contractible. We
discuss possible quantum ground states up to Hopf charge $Q=7$.
\end{abstract}

\maketitle

\section{Introduction}
\label{sec:intro}
\news

The possibility of knot-like solitons in a nonlinear field theory was
first proposed by Faddeev in 1975, \cite{Faddeev:1975}. In 1997, 
interest in the model was revived by an article by Faddeev and Niemi
\cite{Faddeev:1997zj}: the advent of larger computer power and a better
understanding of the initial conditions led to a series of papers. In
\cite{Gladikowski:1997mb} axially symmetric configurations were studied
extensively. Papers by Battye and Sutcliffe showed that for higher Hopf
charge twisted, knotted and linked configurations occur
\cite{Battye:1998pe, Battye:1998zn}. The most recent results are due to
Hietarinta and Salo \cite{Hietarinta:1998kt, Hietarinta:2000ci}. 
Stable and metastable static solutions have now been explored up to Hopf
charge $Q=8$.

Quantization of Hopf solitons was first discussed in 
\cite{Gladikowski:1997mb}. More recently Su described a collective
coordinate quantization in \cite{Su:2001zw} which was motivated by the
collective coordinate quantization of Skyrmions in \cite{Adkins:1983ya}.
However, collective coordinate quantizations can be potentially misleading
unless the topology of configuration space is examined carefully
\cite{Balachandran:1991}.  

In this paper we describe the fermionic quantization of Hopf solitons
following an old idea of Finkelstein and Rubinstein
\cite{Finkelstein:1968hy}. Solitons in scalar field theories can
consistently be quantized as fermions provided the fundamental group of
configuration space has a ${\mathbb Z}_2$ subgroup generated by a loop
in which two identical solitons are exchanged. 
Loops in configuration space give
rise to so-called Finkelstein-Rubinstein constraints which depend on
whether the loop is contractible. The Skyrme model
\cite{Skyrme:1961vq} was the main motivation for this approach; 
see \cite{Krusch:2002by} for further references. Symmetries of
classical configurations induce loops in configuration space. After
quantization these loops give rise to constraints on the wave function.
Recently, a simple formula has been found to determine whether a loop in
the configuration space of Skyrmions is contractible
\cite{Krusch:2002by}. We shall exploit the fact that
 Skyrmions and Hopf solitons are related via the Hopf map to use
Skyrmions as a tool to study Hopf solitons. 

This paper is organized as follows. In section \ref{topology} we discuss the
configuration space of Hopf solitons for general domains. The
configuration space of Skyrmions can be related to Hopf solitons via the
Hopf map which is a fibration. This mathematical structure enables us to
prove that the Hopf map induces, in certain circumstances,
 an isomorphism between the fundamental
groups of the Skyrme and Faddeev-Hopf configuration spaces. 
In section \ref{Hopf} we summarize some known facts about Hopf solitons.
In section \ref{FR} we describe how to quantize a Hopf soliton as a fermion
and calculate possible ground states in the Faddeev-Hopf model.
In the following section, we discuss collective coordinate quantization 
in this context. We end with some concluding remarks.

%
\section{The topology of configuration space}
\label{topology}

Let $M$ be a compact, connected, oriented 3-manifold and $p_0\in M$ be a 
marked point. The case of most interest is $M=S^3$, interpreted
as the one point compactification of $\R^3$ with $p_0$ representing the 
boundary at infinity. The configuration space we seek to study is
$(S^2)^M$, the space of based maps $M\ra S^2$, that is continuous maps
sending the
chosen point $p_0$ to a chosen point in $S^2$, $(0,0,1)$ 
say. We also define the space $\free{M}{S^2}$ of unbased maps $M \ra S^2$
and similarly $(S^3)^M$ and $\free{M}{S^3}$ where the chosen point is
$(1,0)\in S^3\subset\C^2$, say. All such spaces are given the
compact open topology (equivalent to the $C^0$ topology). Our goal in
this section is to relate the topology of $(S^2)^M$, the
Faddeev-Hopf configuration space, to that of $(S^3)^M$, the standard
Skyrme configuration space. 

The connected components of $(S^2)^M$ were enumerated and classified by
Pontrjagin \cite{pon}. Let $\mu$ be a generating 2-cocycle for
$H^2(S^2;\Z)= \Z$. Then given $\phi\in (S^2)^M$ one has an associated
2-cocycle $\phi^*\mu\in H^2(M;\Z)$ by pullback. No two maps
$M\ra S^2$ having noncohomologous 2-cocycles can be homotopic, 
and every 2-cocycle on $M$ 
is cohomologous to
 the pullback of $\mu$
 by some map. Thus, the homotopy classes of maps $M\ra S^2$
fall into disjoint families labelled by $H^2(M;\Z)$. Within any such family,
the classes are labelled by elements of 
$H^3(M;\Z)/2[\phi^*\mu]\cup H^1(M;\Z)$.
Note that this group varies from family to family and that to compute it
requires knowledge of the ring structure on $H^*(M;\Z)$. 
The most important family is the one with $[\phi^*\mu]=0$, the so-called
algebraically inessential maps. Classes within this family are labelled by
elements of $H^3(M;\Z)= \Z$, identified with the Hopf charge $Q$,
which we would like to interpret as the soliton number of the
configuration, that is, the excess of solitons over antisolitons. Let
us denote the space of algebraically inessential maps by
$(S^2)^M_*\subset (S^2)^M$. Note that these sets coincide if $H^2(M;\Z)=0$,
for example, when $M=S^3$. Configurations outside $(S^2)^M_*$ wrap some
2-cycle in $M$ nontrivially around $S^2$. They are bound to some
topological defect in physical
space and so are arguably not localized topological
solitons at all. We shall not consider their physics in this paper.

Our main tool will be the Hopf map $\pi:S^3\ra S^2$, most conveniently
defined by identifying $S^3$ with the unit sphere in $\C^2$ and $S^2$
with $\CP^1$, for then 
\beq
\label{Hopfmap}
\pi:(z_1,z_2)\mapsto [z_1,z_2]. 
\eeq
Note that $\pi$ sends the marked point $(1,0)\in S^3$ to the marked point
$[1,0]\in S^2$, corresponding to the North pole, $(0,0,1)$.
The map $\pi$ is
a fibration, that is, it has the homotopy lifting property with respect to
all domains. A map $\phi:M\ra S^2$ has a lift $\wt{\phi}:M\ra S^3$
(where $\pi\circ\wt{\phi}=\phi$) if and only if $\phi^*\mu =0$, that is,
if and only if $\phi\in (S^2)^M_*$. The integer in $H^3(M;\Z)$ labelling
the class of $\phi$ is precisely the degree of $\wt{\phi}:M\ra S^3$,
that is, the baryon number of the Skyrme configuration $\wt{\phi}$. 
This was shown explicitly for $M=S^3$ in \cite{Meissner:1985}. So, given a
Skyrme
configuration of degree $Q$, we may produce an algebraically inessential
Hopf configuration of charge $Q$ by composition with the Hopf map. In this 
way we produce a map $\pi_*:(S^3)^M\ra (S^2)^M_*$. To what extent does the
topology of $(S^3)^M$ determine that of $(S^2)^M_*$? 

\begin{thm} The map $\pi_*:(S^3)^M\ra (S^2)^M_*$ induced by the Hopf 
fibration is a Serre fibration.
\end{thm}

\noindent
{\it Proof:} We must prove that
the map has the homotopy lifting property with
respect to all {\em disks} $D^k$ \cite{serre}, that is, that the commutative
diagram below left  may be completed by a map $\wt{H}$ along the diagonal.
Here $H$ is a homotopy between two maps $f_0,f_1:D^k\ra (S^2)^M_*$ and
$\wt{f}_0$ is a lift of $f_0$. Using the identification of $g:X\ra Y^Z$
with $\hat{g}:Z \times X \ra Y$, we produce the commutative 
diagram below right. Now the homotopy $\hat{H}$ certainly does lift 
to $\wt{\hat{H}}$ since
$\pi$ is a fibration. From $\wt{\hat{H}}$ we produce a map 
$\check{H}:D^k\times I\ra
\free{M}{S^3}$ by $(\check{H}(d,t))(p)=\wt{\hat{H}}(p,d,t)$. A priori, this 
is not necessarily the lifted homotopy we seek, however, since there is no 
reason why it should respect the basing condition. 

\begin{center}
\begin{tabular}{ccc}
\begin{diagram}
D^k\times \{0\} & \rTo^{\wt{f}_0} & (S^3)^M   \\
\dTo^{\iota} & \ruTo^{\wt{H}} & \dTo_{\pi_*} \\
D^k\times [0, 1] & \rTo_{H} & (S^2)^M_*\\
\end{diagram}
&\hspace{1cm}&
\begin{diagram}
M\times D^k\times \{0\} & \rTo^{\hat{\wt{f}}_0} & S^3   \\
\dTo^{\iota} & \ruTo^{\wt{\hat{H}}} & \dTo_{\pi} \\
M\times D^k\times [0, 1] & \rTo_{\hat{H}} & S^2\\
\end{diagram}
\end{tabular}
\end{center}

Let $U\subset
S^2$
be a small closed ball centred on $(0,0,1)$ and choose a local trivialization
of the Hopf bundle $S^1\hookrightarrow S^3\stackrel{\pi}{\ra} S^2$ over $U$.
 Then by
continuity of $\hat{H}$ and compactness of $D^k\times [0,1]$,
there exists a closed ball $B\subset M$ centred
on $p_0$ so that the restriction $\wt{\hat{H}}|:B\times D^k\times I\ra
S^3$ takes values in $\pi^{-1}(U)$. We may write it, with respect to
our local trivialization, as
$$
\wt{\hat{H}}|(p,d,t)=(\hat{H}(p,d,t),\lambda(p,d,t))
$$
where $\lambda:B\times D^k\times I\ra S^1$. In this language, we are
done if $\lambda|:\{p_0\}\times D^k\times I\ra \{1\}$, for then the map
$\check{H}$ {\em does} satisfy the basing criteria. Note that we are
free to change $\lambda$ to any continuous map $\lambda_*$
 we please, provided we
do not change it on $\cd B\times D\times I$, since this just shifts
$\wt{\hat{H}}$ along the fibres of $S^3$ which does not change 
$\pi\circ\wt{\hat{H}}$, so that the altered map is still a lift of 
$\hat{H}$,
and is still continuous. 
Now since $\cd B\times D^k\times I$ deformation retracts to
$S^2$ and $\pi_2(S^1)=0$, $\lambda|:\cd B
\times D^k\times I$ is nullhomotopic and we may construct the
required $\lambda_*:B\times D\times I\ra S^1$ by applying the null
homotopy radially in $B$.\hfill $\Box$

Our main interest is to understand the fundamental
group of each connected component of $(S^2)^M_*$. Given any map 
$\rho:X\ra Y$, 
there is a natural homomorphism $\rho_*:\pi_1(X)\ra \pi_1(Y)$
defined by composition of loops in $X$ with $\rho$. The fact
that $\pi_*$, which we will henceforth denote $\rho$, 
is a Serre fibration allows us to 
obtain a short exact sequence relating $\pi_1((S^3)^M)$ and
$\pi_1((S^2)^M_*)$. In the case $M=S^3$ this reduces to the statement that
the
homomorphism $\rho_{*}$ associated with $\rho$ 
is actually an isomorphism. We can therefore determine the
homotopy class of a loop in the Hopf configuration space by lifting it
to a loop in the Skyrme configuration space and applying known results.

\begin{thm} The map $\rho:(S^3)^M\ra (S^2)^M_*$ obtained from the Hopf 
fibration induces a short exact
sequence of groups
$$
0\ra \pi_1((S^3)^M)\stackrel{\rho_*}{\ra}
 \pi_1((S^2)^M_*)\ra H^1(M;\Z)\ra 0.
$$
\end{thm}

\noindent
{\it Proof:}
Given any Serre fibration $F\hookrightarrow E\stackrel{\rho}{\ra} B$,
where $F,E,B$ denote the fibre, total space and base,
we have an induced long exact sequence of homotopy groups:
\bea
\label{les1}
\ldots\ra
\pi_1(F){\ra}\pi_1(E)\stackrel{\rho_*}{\ra}\pi_1(B)\ra\pi_0(F)\ra
\pi_0(E)\stackrel{\rho_*}{\ra}\pi_0(B)\ra 0.
\eea
In the case at hand, $E=(S^3)^M$, $B=(S^2)^M_*$ and $F=(S^1)^M$. Using
the identification $S^1= U(1)$, we see that
$F$ is a topological group, so all its connected components are homeomorphic.
The components of $G^M$ for any Lie group $G$ are enumerated in
\cite{Auckly:2002hb}
while $\pi_1(G^M)$ is constructed in \cite{aucspe}.
 The relevant results here are $\pi_0(F)= H^1(M;\Z)$
and $\pi_1(F)=0$. Note also that $\pi_0(E)=\pi_0(B)= H^3(M;\Z)
=\Z$ by the theorems of Hopf and Pontrjagin.
Substituting in (\ref{les1}) gives
\beq\label{les2}
0\ra \pi_1(E)\stackrel{\rho_*}{\ra}\pi_1(B)\ra H^1(M;\Z)\ra 
\Z\stackrel{\rho_*}{\ra}\Z\ra 0.
\eeq
By exactness, the second $\rho_*$ is surjective, and there are only two
surjective homomorphisms
 $\Z\ra \Z$ (namely $1\mapsto 1$ and $1\mapsto -1$), both 
of which are injective. So we see that the second $\rho_*$ is an isomorphism.
Since the second $\rho_*$ has trivial kernel, the image of $H^1(M)$ in
$\Z$ is $0$ by exactness, and the sequence truncates as was claimed.
\hfill $\Box$

We note in passing that this provides an algebraic proof that the Hopf
map takes degree $Q$ Skyrme configurations to Hopf charge $Q$ (or
$-Q$ if the orientation on $M$ or $S^3$ is swapped) Faddeev-Hopf
configurations, since this is precisely the statement that
$\rho_*:\pi_0((S^3)^M)\ra \pi_0((S^2)^M_*)$ is an isomorphism. 
By identifying the Hopf degree of $\phi\in(S^2)^M_*$ with the degree of 
its lift
$\wt{\phi}\in (S^3)^M$, we adopt the standard
convention that the Hopf map $\pi\in (S^2)^{S^3}$ itself has
Hopf degree $+1$.

The
short exact sequence does not tell us precisely what $\pi_1((S^2)^M_*)$ is
in  general. One useful class of domains (which includes $M=S^3$) where
we do know the answer is those with finite fundamental group.

\begin{cor} 
\label{2}
If $\pi_1(M)$ is finite then $\rho_*:\pi_1((S^3)^M)\ra
\pi_1((S^2)^M_*)$ induced by the Hopf map is an isomorphism.
\end{cor}

\noindent {\it Proof:}
The result follows once we show that $H^1(M;\Z)=0$. By the Universal 
Coefficient Theorem, $H^1(M;\Z)$ is isomorphic to the free part of
$H_1(M;\Z)$, since $H_0(M;\Z)=\Z$ has no torsion. But $H_1(M;\Z)$ is 
isomorphic to the abelianization of $\pi_1(M)$ which, being finite, can have
no free part.\hfill $\Box$

These results are useful because
a lot is known about the topology of $(S^3)^M$
since it can be identified with the topological group
$G^M$ where $G=SU(2)$. The canonical identification is given by
\begin{equation}
\label{SU(2)id}
S^3 \to SU(2): (z_1,z_2) \mapsto 
U= \left(
\begin{array}{cc}
z_1 & - {\bar z_2} \\
z_2 & {\bar z_1} 
\end{array}
\right).
\end{equation}
This map is well-defined because $U^\dagger U = U U^\dagger = \I_2$ and 
$|z_1|^2 + |z_2|^2 = 1$ implies that $\det U = 1$. Also note that 
$(1,0) \mapsto \I_2$.
Since $(SU(2))^M$ is a topological group all connected components of
$(S^3)^M$ are homeomorphic, and the fundamental group is abelian. 
A loop in the
identity component of $SU(2)^M$ based at the constant map $M\ra \{\I_2\}$
may be thought of as a map from $S^1\wedge M$ to $SU(2)$,
where $\wedge$ denotes smash product. If $M=S^3$ then
$S^1\wedge M= S^4$ and
$\pi_4(SU(2))= \Z_2$, so we have that $\pi_1((S^2)^{S^3}_*)=
\pi_1(SU(2)^{S^3})= \Z_2$ for all components. 
Using a similar argument for the {\em vacuum sector} $(S^2)^M_0$ of
the Faddeev-Hopf model, we could very easily have shown that, for $M=S^3$,
$\pi_1((S^2)^M_0)=\pi_4(S^2)= \Z_2$. Note that we have actually
proved much more than this, however: the fundamental group of {\em every}
connected component of the Faddeev-Hopf configuration space is $\Z_2$, and
crucially, that the map from the
Skyrme configuration space induced by Hopf fibration is an isomorphism.

The above results will suffice for our purposes. 
In fact, one can say much more about the algebraic topology of $(S^2)^M$,
with $M$ a general compact oriented 3-manifold.
It turns out that all components of
$(S^2)^M_*$ are homeomorphic, though the same fails
to be true for the full space $(S^2)^M$. Furthermore, it is possible to
compute both the fundamental group and the whole real cohomology ring 
(including its cup product structure) of any
component of $(S^2)^M$. These results are obtained
\cite{aucspe} by exploiting a somewhat less obvious relationship
between $(S^2)^M$ and the vacuum (degree 0) sector of $SU(2)^M$. 
Essentially, all Faddeev-Hopf configurations in a given sector may be
obtained from a fixed map in that sector by acting on the codomain 
with some degree 0 Skyrme
configuration. This gives natural maps from the vacuum sector of the
Skyrme model to each sector of the Faddeev-Hopf model, which can be shown to
have many topologically natural properties. 
The topological
results we present here are not so powerful as those of \cite{aucspe},
but they are also less technical and may be
visualized rather concretely. Most importantly, they are particularly
well-suited to the study of Finkelstein-Rubinstein quantization in the
Faddeev-Hopf model. 


\section{The Faddeev-Hopf model}
\news
\label{Hopf}

From now on we consider only the case
$M = S^3$, interpreted as the one point
compactification of ${\mathbb R}^3$ with the point $p_0$ representing the
boundary at infinity. The most extensively studied model of this kind is
due to Faddeev \cite{Faddeev:1975} who suggested the following
Lagrangian density
\begin{equation}
\label{Lagrangian}
{\cal L} = \frac{1}{2}\partial_\mu \nvec \cdot \partial^\mu \nvec  -
\frac{\lambda}{4} \left( \partial_\mu \nvec \times \partial_\nu \nvec \right)
\cdot  \left(\partial^\mu \nvec \times \partial^\nu \nvec \right)
\end{equation}
where the field $\nvec = (n_1,n_2,n_3)$  takes values on the 2-sphere, that
is $|\nvec|^2 = 1$, $\lambda$ is a coupling constant,
and the boundary condition is $\nvec(\infty) = (0,0,1)$.
We have changed notation from $\phi$ to $\nvec$ for the field so as to fit
in with the existing literature on the model.
Note that the second term in (\ref{Lagrangian}) stabilizes the solitons
against radial rescaling. As discussed in section \ref{topology} the
Hopf charge $Q$ can be identified with the degree of any lift of $\nvec$ to
$\wt{\nvec}:\R^3\ra S^3$. 
The energy $E$ of a static configuration of Hopf charge $Q$ is bounded
below by
\beq
E \ge c |Q|^\frac{3}{4}
\eeq
where $c$ is a constant. For more details see \cite{Vakulenko:1979uw,
Ward:1998pj}.

The Lagrangian of the model has $E(3) \times O(3)$ symmetry.
Since spatial translations are rather trivial we will not discuss them any
further. The target space $O(3)$ symmetry is broken to $O(2)$ symmetry by
the boundary condition. 
Kundu and Rybakov showed in \cite{Kundu:1982bc} that
topologically nontrivial configurations admit at most an axial
(one-parameter) symmetry. General configurations with axial symmetry are
discussed in \cite{Gladikowski:1997mb}.
Special configurations with axial symmetry have been studied recently
in \cite{Hietarinta:2000ci} and can be described in the following way.
Introduce toroidal coordinates $(\eta, \xi, \phi)$ on $\R^3$ defined by
\beq
x = a \frac{\sinh \eta \cos \phi}{\cosh \eta - \cos \xi},~
y = a \frac{\sinh \eta \sin \phi}{\cosh \eta - \cos \xi},~
z = a \frac{\sin \xi}{\cosh \eta - \cos \xi}.
\eeq
These coordinates form a canonically oriented orthogonal system
covering all of $\R^3$ except the circle $C=
\{x^2+y^2=a^2,z=0\}$
and the $z$-axis. Surfaces of constant $\eta\in (0,\infty)$ are tori
of revolution about the $z$-axis, but with non-circular generating curves.
As $\eta\ra\infty$ these tori collapse to the circle $C$ and as $\eta\ra 0$
they collapse to the $z$-axis. Each torus of constant $\eta$ is parametrized
by the angular coordinates $(\phi,\xi)$; 
$\phi$ is the angle around the $z$ axis, 
$\xi$ is an angular coordinate around the not quite circular generating
curve of the torus. The maps of interest are most easily written in terms
of a complex stereographic coordinate $W$ on $S^2$. Projecting from
$(0,0,1)$, so that  $W=(n_1+in_2)/(1-n_3)$,
they take the form\footnote{Note that we have
changed the sign of $n$ in \cite{Hietarinta:2000ci}.} 
\beq
\label{Utorus}
W = f(\eta) {\rm e}^{i(m \xi - n \phi)},
\eeq
where $f(\eta)$ satisfies the boundary conditions $f(0) = \infty$ and
$f(\infty) = 0$. Inverting the stereographic projection yields
\beq
\label{ntorus}
\nvec 
= \left( \frac{2f}{f^2+1} \cos(m \xi - n \phi),
\frac{2f}{f^2+1} \sin(m \xi - n \phi),
\frac{f^2-1}{f^2+1}
\right).
\eeq
This ansatz will be referred to as the toroidal ansatz. Here the
word ``ansatz'' is used rather loosely, for an approximation which is a
good initial guess for the numerically calculated static solution.
It is worth mentioning that the toroidal ansatz
gives rise to exact solutions for the Lagrangian density
$
{\cal L} = (H_{\mu \nu}H^{\mu\nu})^\frac{3}{4}
$
where $H_{\mu \nu} = \nvec \cdot(\partial_\mu \nvec \times \partial_\nu
\nvec)$, \cite{Aratyn:1999cf}.

Under rotation by $\alpha$ around the $z$ axis the toroidal coordinates
change to $(\eta, \xi, \phi + \alpha)$ which rotates the vector $\nvec$ by
$-n \alpha$ around the third axis in target space. Obviously, this
rotation can be undone by a rotation around the third axis in target
space.

There is an obvious lift of any map $\R^3\ra S^2$ within this ansatz to
a Skyrme configuration $\R^3\ra S^3$, obtained as follows.
For given $f$, $m$ and $n$, let 
\beq
\label{1}
\wt{\nvec}:(\eta,\phi,\xi)\mapsto (z_1,z_2)\in\C^2,\qquad\mbox{where}\qquad
z_1 =  \frac{f}{\sqrt{f^2+1}}{\rm e}^{i m \xi}, \quad
z_2 =  \frac{1}{\sqrt{f^2+1}}{\rm e}^{i n \phi}.
\eeq
Then $|z_1|^2 + |z_2|^2 = 1$ so that 
$\wt{\nvec}$ is actually $S^3$-valued, 
and the composition of this map with the Hopf
map is clearly $\nvec$, since the stereographic coordinate $W$ coincides
with the inhomogeneous coordinate $W=z_1/z_2$ under the identification
$S^2\equiv\CP^1$. It is now straightforward to compute the degree
of $\wt{\nvec}$, and hence the Hopf degree of $\nvec$. Since the degree
of $\wt{\nvec}$ is a homotopy invariant, we may deform $f$ to any convenient
function satisfying the boundary conditions, for example, 
$f(\eta)=\eta^{-1}$. In this case, $(2^{-\frac{1}{2}},2^{-\frac{1}{2}})\in
S^3$ is a regular value of $\wt{\nvec}$ with precisely $|mn|$ preimages,
namely the points with $\eta=\exp(im\xi)=\exp(in\phi)=1$. At each of
these preimages, the image of  the canonically oriented 
coordinate frame under $d\wt{\nvec}$ is
$$
d\wt{\nvec}:[\cd_\eta,\cd_\xi,\cd_\phi]\mapsto
[(-2^{-\frac{3}{2}},0,2^{-\frac{3}{2}},0),(0,\frac{m}{\sqrt{2}},0,0),
(0,0,0,\frac{n}{\sqrt{2}})]
$$
where we have identified $\C^2\cong\R^4$. The orientation of the
image frame is given by the sign of the determinant
$$
\det[d\wt{\nvec}\cd_\eta,\, d\wt{\nvec}\cd_\xi,\, d\wt{\nvec}\cd_\phi,\,
\wt{\nvec}]
=\frac{mn}{8}.
$$
Hence each of the $|mn|$ preimages
 has multiplicity $+1$ if $mn>0$ and $-1$ if $mn<0$,
so the Hopf charge of $\nvec$ is $mn$, in agreement with the calculation in
\cite{Gladikowski:1997mb}. 

Numerical evidence suggests that the energy minimals
for $Q=1,2$ and $4$ have axial symmetry. In general, minimals are
more complicated, having knotted or linked structures with
 at most discrete symmetries. In principle any cyclic group $C_q$
is a possible discrete symmetry. However, in practice only the simplest
 nontrivial symmetry --- the twofold symmetry $C_2$ --- seems to
occur. Clearly any nonconstant smooth field configuration cannot be
symmetric under a rotation in target space without a compensating
spatial rotation. It is possible, however, for a configuration to
be invariant under a spatial rotation without a compensating
rotation in target
space. For example, 
the axial configuration in (\ref{ntorus}) with even $n$ has a $C_2$ symmetry
generated by spatial rotation by $\pi$ about the $z$-axis.
We will discuss symmetries further in the next section when
we calculate the constraints they impose on the wave function.


\section{Finkelstein-Rubinstein constraints}
\news
\label{FR}

In this section we describe how to use ideas of Finkelstein and Rubinstein
\cite{Finkelstein:1968hy} to quantize a scalar field theory and
obtain fermions. Quantization
usually implies replacing the classical configuration space by wave
functions on configuration space. However, if the configuration space
is not simply connected it is possible to define wave functions on the
universal cover of configuration space. As shown in section \ref{topology},
the fundamental group of each connected component of
our configuration space $\Q=(S^2)^{S^3}_*$ 
is
${\mathbb Z}_2$. So the universal cover $\wt{\Q}$ 
is a
twofold cover. We will also assume
that the topological charge is conserved in the quantum theory, as it
is in the classical theory,  so
the wave functions are defined on the covering space of a component
of configuration space $\Q_Q$ with fixed Hopf charge $Q$. We shall formally
think of the quantum state of the model as being specified by a wave function
$\Psi\in L^2(\wt{\Q}_Q)$ with respect to some measure on $\wt{\Q}_Q$.
Let $\Pi:\wt{\Q}_Q\ra\wt{\Q}_Q$ be the deck transformation, that is, the map
  which takes $p$ to the unique
point in $\wt{\Q}_Q$ which differs from $p$ but projects to the same point in
$\Q_Q$. 
This induces a linear map $\Pi^*:L^2\ra L^2$ by pullback:
$(\Pi^*\Psi)(p):=\Psi(\Pi(p))$. Since the states $\Pi^*\Psi$ and $\Psi$
are physically indistinguishable, we must have 
$\Psi(p)=e^{i\theta(p)}\Pi^*\Psi(p)$
for all $p\in\wt{\Q}_Q$ and all $\Psi$. 
But $\Pi^*\Pi^*=1$, so the only possibilities are
$\Pi^*\Psi=\Psi$ or $\Pi^*\Psi=-\Psi$. In order to allow for 
 fermionic solitons, we must consistently choose the latter
possibility: our wavefunctions must always be {\em odd} under $\Pi$. 

Spinoriality then arises as follows. Consider the loop in $\Q_Q$ defined
by spatial rotation about a fixed axis through $2\pi$ of a fixed base
configuration $\nvec$. 
Since $\pi_1(\Q_Q)=\Z_2$, this may not be 
contractible, and its contractibility is independent of the basepoint
$\nvec$ chosen. If it is noncontractible, 
both lifts of the loop to $\wt{\Q}_Q$ fail to
close, but are rather paths connecting a $\Pi$-related pair of points
(both of which project to $\nvec$). Having insisted on $\Pi$-oddness,
therefore, we see that every allowable state in this sector aquires a minus
sign under spatial rotation by $2\pi$, the hallmark of spinoriality. That
this is equivalent to fermionicity (that is, odd exchange statistics)
was proved by Finkelstein and Rubinstein in \cite{Finkelstein:1968hy}.

The question of whether Hopf solitons can be consistently quantized as
fermions thus reduces to the question of whether $2\pi$ spatial rotation
loops in $\Q_Q$ are noncontractible when $Q$ is odd and contractible when
$Q$ is even. To answer this, we only need to determine the contractibility 
for a representative of each sector.
Consider the loop $\gamma:[0,1]\ra (S^3)^{S^3}$
defined by 
$\hat{\gamma}(\eta,\xi,\phi,t)=\wt{\nvec}(\eta,\xi,\phi+2\pi t)$,
where $\wt{\nvec}:S^3\ra S^3$ is defined in (\ref{1}), and we once again
use the natural identification of $g:X\ra Y^Z$
with $\hat{g}:Z\times X\ra Y$. This is a $2\pi$ spatial rotation loop
(about the $z$ axis)
of the degree $Q=mn$ Skyrme configuration $\wt{\nvec}$. Note that
$\pi\circ\gamma:[0,1]\ra\Q_Q$ is also a $2\pi$ rotation loop, but in the
Faddeev-Hopf configuration space. Corollary \ref{2} states that $\pi\circ
\gamma$ is contractible if and only if $\gamma$ is contractible, which is
true if and only if the degree $Q$ is odd, by work of Giulini 
\cite{Giulini:1993gd}.
Hence imposing $\Pi$-oddness on our quantum states $\Psi$ does indeed
produce a consistent fermionic quantization of Hopf solitons.

It is important to realize that, having imposed $\Pi$-oddness, {\em every}
noncontractible loop in $\Q_Q$ must be associated with a sign flip in
$\Psi:\wt{\Q}_Q\ra\C$, regardless of whether the loop is generated by
a spatial rotation. Let $\nvec$ be a Hopf degree $Q\neq 0$ energy minimal
of the Faddeev-Hopf model which is invariant under a simultaneous
spatial rotation by $\alpha$ about some axis $\evec$ and rotation by 
$\beta$ around the third axis in target space (the only axis 
compatible with the boundary conditions). Since for $Q\neq 0$ the
maximal symmetry of a configuration is $O(2)\times O(2)$, only one spatial
rotation axis $\evec$
is possible for a given $\nvec$, and we may choose it, without
loss of generality, to lie along the $z$ axis. Let us call such a
combined transformation an $(\alpha,\beta)$-rotation.
Then we may construct a loop
$L(\alpha, \beta)_{\nvec}$ in $\Q_Q$ based at $\nvec$ which consists of
rotation by $2t\alpha$ around the $z$-axis for time 
$t\in [0,\frac{1}{2}]$, followed by  rotation by 
$(2t-1)\beta$
around the third axis in target space 
for $t\in(\frac{1}{2},1]$. In this language, the fact that $\nvec$ has the 
specified symmetry is precisely the statement that $L(\alpha,\beta)_\nvec$
is a loop, i.e.\ closed. There are two points $p,\Pi(p)\in\wt{\Q}_Q$
corresponding to $\nvec$, and any physical state must have $\Psi(\Pi(p))=
-\Psi(p)$. 
Now if $L(\alpha,\beta)_\nvec$ is noncontractible
then $p$ and $\Pi(p)$ are connected by the lifts of $L(\alpha,\beta)_\nvec$,
starting at $p$ and $\Pi(p)$, respectively.
Hence, evaluated at the specific point $p$ (or $\Pi(p)$) we must have
\beq
\label{3}
({\rm e}^{-i \alpha {\hat L}_3} 
{\rm e}^{-i \beta {\hat K}_3} \Psi)(p)=-\Psi(p),
\eeq
 for any allowed state, where 
${\hat L}_3$ is the third component of the spin operator $\hat{\bf L}$
and ${\hat K}_3$ is the third (and only) component
of the spin operator in target space (henceforth called
isospin).

If $L(\alpha,\beta)_\nvec$ were contractible, however, it would lift to a
pair of closed loops in $\wt{\Q}_Q$ based at $p$ and $\Pi(p)$, so that
\beq
\label{4}
({\rm e}^{-i \alpha {\hat L}_3} 
{\rm e}^{-i \beta {\hat K}_3} \Psi)(p)=\Psi(p),
\eeq
simply by continuity of $\Psi$.
 In the spirit of semiclassical
quantization we assume that, at least for low lying states,
 the symmetry of the classical energy minimal
is not broken by quantum effects. Thus we seek quantum states $\Psi$
which are also invariant under $(\alpha,\beta)$-rotations, so that
\beq
({\rm e}^{-i \alpha {\hat L}_3} 
{\rm e}^{-i \beta {\hat K}_3} \Psi)(x)=e^{i\theta(x)}\Psi(x),
\eeq
for all $x\in\wt{\Q}_Q$. But, assuming the $(\alpha,\beta)$-rotation 
generates a finite group, there must exist an integer $q$ such that
$({\rm e}^{-i \alpha {\hat J}_3} 
{\rm e}^{-i \beta {\hat I}_3})^q\Psi\equiv\Psi$,
 which implies, by continuity,
that $\theta(x)$ must in fact be constant. But then $\theta(x)=\theta(p)=\pi$
if $L(\alpha,\beta)_\nvec$ is noncontractible
by (\ref{3}), or $\theta(x)\equiv 0$ if $L(\alpha,\beta)_\nvec$ is
contractible, by (\ref{4}). Hence, we obtain the
so-called Finkelstein-Rubinstein constraints on symmetric quantum
states:
\beq
\label{FRconstraint}
{\rm e}^{-i \alpha {\hat L}_3} 
{\rm e}^{-i \beta {\hat K}_3} \psi = 
\left\{
\begin{array}{rcl}
\psi & ~~~ & {\rm if~the~induced~loop~is~contractible,} \\
-\psi & ~~~ & {\rm otherwise.}
\end{array}
\right.
\eeq
Equation (\ref{FRconstraint}) imposes constraints on the spin
and isospin quantum numbers $L, L_3$ and $K_3$.

It is worth pausing here to discuss the relationship between
body-fixed and space-fixed angular momentum. The Lagrangian of the
Hopf model is invariant under a $SO(3) \times SO(3)$ symmetry group
consisting of rotations in space and target space. For these symmetries
we can define left and right actions which are generated by the
space-fixed and body-fixed angular momenta ${\bf J}$ and ${\bf L}$ 
acting on space and by space-fixed and body-fixed angular momenta 
${\bf I}$ and ${\bf K}$ acting on target space. The body-fixed and
space-fixed angular momentum operators are related by rotations which
implies that  ${\bf J}^2 = {\bf L}^2$. For rotations in
target space only rotations around the third axis are compatible with
the boundary conditions. This implies $I_3^2 = K_3^2$.
When the model is quantized the angular momentum operators 
${\hat {\bf J}}^2={\hat {\bf L}}^2$, ${\hat J}_3$, ${\hat L}_3$,
${\hat I}_3$ and ${\hat K}_3$ form a set of commuting observables. The
quantum wave function $\psi$ can then be labelled by the usual spin quantum
number as follows $\psi = | L, L_3, J_3, K_3, I_3 \rangle.$ Since the
Finkelstein-Rubinstein constraints do not impose any restrictions on
the values of $J_3$ and $I_3$, these values will often be suppressed
and the wave function is given as $\psi = | L, L_3, K_3 \rangle$. 
%
In order to make predictions, we are interested in states with given
$J$ and $I_3$. Therefore, we have to consider states with quantum
numbers $L=J$ and $K_3 = \pm I_3$. Then the Finkelstein-Rubinstein
constraints have the following effect. By 
restricting the allowed quantum states for given $J$ and $I_3$ 
the degeneracy of the states is changed. 
In the extreme case that the degeneracy is zero, certain combinations
of $J$ and $I_3$ get excluded.

We now return to our discussion of loops in configuration space and
Finkelstein-Rubinstein constraints.
Just as for $2\pi$ spatial rotation loops, we can use the isomorphism
$\pi_1((S^2)^{S^3}_*)\ra\pi_1((S^3)^{S^3})$ induced by the Hopf fibration
to calculate whether a given loop $L(\alpha,\beta)_\nvec$ is 
contractible. For every
configuration $\nvec$ we can choose a configuration ${\tilde \nvec}$ in the
configuration space $(S^3)^{S^3}$ of Skyrmions. Then 
$L(\alpha, \beta)_{\tilde \nvec}$ is
a loop in $(S^3)^{S^3}$ which projects to the loop 
$L(\alpha, \beta)_\nvec$ in $(S^2)^{S^3}_*$ under $\pi$. 
The action of $SO(3)$ on the target space of $\wt{\nvec}$, that is $S^3$,
is now identified with the adjoint action of $SU(2)$ on itself.
Once again, 
Corollary \ref{2} shows that $L(\alpha, \beta)_\nvec$
is contractible if and only if $L(\alpha, \beta)_{\tilde{\nvec}}$ is
contractible. Contractibility of the latter loop can be determined by
means of an explicit formula recently derived for Skyrmions with
discrete symmetries, \cite{Krusch:2002by}. This states that the loop 
$L(\alpha, \beta)_{\tilde \nvec}$ is contractible if and only if
\beq
\label{N}
N = \frac{Q}{2 \pi} \left(Q \alpha - \beta \right)
\eeq
is even. Note that there is a slight subtlety with the choice of the sign
of $\beta$. 

We can immediately recover our earlier result that the $\Pi$-odd quantization
is consistently fermionic from formula (\ref{N}). To see this, note that
every configuration is symmetric under $(2\pi,0)$-rotation, and substituting
$\alpha=2\pi$, $\beta=0$ into (\ref{N}) shows that $N$ is odd if and only if
$Q$ is odd. Hence
the spin quantum numbers $L$ and $J$ are half integer
if and only if $Q$ is odd. Similarly, considering the case $\alpha=0$, 
$\beta=2\pi$ (pure isorotation by $2\pi$) shows 
that
the isospin quantum numbers $K_3$ and $I_3$ are also half integer if
and only if $Q$ is odd. 

New constraints on low-lying quantum states $\Psi$ are obtained if we assume
that they are invariant under the symmetry groups of the corresponding 
classical
energy minimals. The Faddeev-Hopf model has received much less numerical
attention than the Skyrme model, so our understanding of these minimals
and their symmetries is comparatively limited. For this reason, we will
discuss the Finkelstein-Rubinstein constraints for general symmetries
first, then apply the analysis to those symmetries which have been
observed in numerical experiments.
Since we are interested in symmetries which can be generated by loops in
configuration space we disregard reflections and look only at
subgroups of $T^2=SO(2)\times SO(2)$. Note that $T^2$, and hence every
subgroup of $T^2$, is abelian. This severely limits the symmetry groups
possible, and accounts in part for the numerical observation that
Hopf degree $Q$ minimals tend to possess far less symmetry than degree
$Q$ Skyrmions. The symmetry group $G_\nvec<T^2$ of a configuration $\nvec$
is either continuous, in which case $G_\nvec\cong SO(2)$
corresponding to axial symmetry, or discrete, hence finite ($T^2$ is
compact). Every finite abelian group is isomorphic to a product of
finite cyclic groups of coprime order, so it suffices to understand
the Finkelstein-Rubinstein constraints for $q$-fold cyclic symmetry
$C_q$. 

First, we deal with axial symmetry. 
Consider the axial configurations (\ref{Utorus}) with Hopf
charge $Q = mn$. These are invariant under $(\alpha,n\alpha)$-rotations
for all $\alpha\in\R$.
 Since the loop $L(\alpha, n \alpha)_\nvec$ exists  for all $\alpha \in
{\mathbb R}$ it is homotopic to the constant loop ($\alpha=0$). 
So $L(\alpha, n \alpha)_\nvec$ is contractible and gives rise to the 
following constraint on wave functions:
\beq
\label{axial}
{\rm e}^{-i \alpha {\hat L}_3} 
{\rm e}^{-i n \alpha {\hat K}_3} \Psi = \Psi.
\eeq
Since formula (\ref{axial}) is valid for all $\alpha$ we can expand the
equation in $\alpha$. The first order term gives rise to the following
constraint for the spin operators:
\beq
\label{Caxial}
({\hat L}_3 + n {\hat K}_3) \Psi = 0.
\eeq 
Equation (\ref{Caxial}) implies for the spin quantum numbers 
that $L_3 = - n K_3$.

If the axial symmetry is broken then the symmetry group must be isomorphic to
a product of finite cyclic groups. Not every cyclic subgroup of $T^2$
is possible for a given $Q$, however, since the generator
$(\alpha,\beta)$ of $C_q<T^2$ must satisfy equation (\ref{N}), that is,
$N$ must be an {\em integer}. There are precisely $q$ different $C_q$
subgroups of $T^2$ which are candidates for symmetry groups,  generated
by $(2\pi/q,2k\pi/q)$ where $k=0,1,\ldots,q-1$, since
pure isorotation can never leave a nonconstant configuration invariant.
Let us denote these groups $C_q^k$. 
To illustrate, let us assume that $q$ is prime
so that $C_q$ is a finite field. Then formula (\ref{N})
 applied to the generator of
$C_q^k$ implies that $Q(Q-k)= 0\mod q$ and hence
 $Q=0 \mod q$ or $Q=k \mod q$ by the field property.
Hence, unless $Q$ is a multiple of $q$, formula (\ref{N})
rules out all possible $C_q$ symmetries except $C_q^{Q\mod q}$. Similar
criteria can be derived for $q$ not prime, but they are not so neat.
Of particular interest given the current state of numerics is the case
$q=2$. The argument above shows that, for odd $Q$, only $C_2^1$ symmetry
is possible, not $C_2^0$. 

Given a candidate symmetry group $C_q^k$, formula (\ref{N}) gives us a
one-dimensional (hence irreducible) representation of $C_{\bar{q}}$, where
$\bar{q}=q$ if $Q(k+1)$ is even and $\bar{q}=2q$ if
$Q(k+1)$ is odd, by mapping the
generator $(2\pi/q,2k\pi/q)$ to $(-1)^N$. This representation
may also be thought of as a homomorphism ${C}_{\bar{q}}
\ra\Z_2=\{1,-1\}$ and is thus
necessarily trivial if $q$ is odd and $Q(k+1)$ is even. 
We call this the
Finkelstein-Rubinstein representation 
of ${C}_{\bar{q}}$. 
There is also a natural representation of ${C}_{\bar{q}}$ on the
spin-isospin $L,K_3$ quantum state space, defined by the
inclusion ${C}_q^k <  SO(3)\times SO(2)$. A state $\Psi$ with quantum
numbers $L,K_3$ is thus compatible with $C_q^k$ symmetry if
and only if
the decomposition of the spin-isospin $L,K_3$
representation of ${C}_{\bar{q}}$ into
irreducible representations contains
a copy of the Finkelstein-Rubinstein representation. 
Given that we consider only cyclic groups,
in practice we need only check compatibility on the generator
$(\alpha,\beta)=(2\pi/q,2\pi k/q)$. Thus
$L_3,K_3$ must satisfy
\bea
e^{-2\pi i(L_3+k K_3)/q}&=&(-1)^N=e^{i\pi Q(Q-k)/q}\\
\Leftrightarrow\qquad L_3+k K_3&=&-\frac{1}{2}Q(Q-k)+\ell q
\eea
where $\ell$ is an integer.

A good candidate for the ground state in the charge $Q$ sector
is the state with
the lowest values of $L$ and $|K_3|$ (and hence $J$ and $|I_3|$)
compatible in this way with the symmetries of the classical minimal.

To illustrate this symmetry analysis, we compute the quantum ground state
for stable and metastable Hopf solitons of degrees $Q=1,\ldots 7$, using 
the classical solutions obtained numerically by Hietarinta et al
\cite{Hietarinta:2000ci}. Only axial and $C_2$ symmetries ever arise for
these solutions. 
In the $C_2$ case for {\em even} $Q$, we distinguish between the two possible
groups $C_2^0$ and $C_2^1$ using the colour coding information in
\cite{Hietarinta:2000ci}.  
The results are presented in table \ref{table1}. 
The first entry is the Hopf number $Q$. A star
indicates that the state is metastable, that is, the classical solution is 
not a global minimal.
The next entry is the energy $E_Q$ which has been calculated in
\cite{Hietarinta:2000ci} and corresponds to $\lambda = 1/4$. The
following entry gives the shape of the Hopf configuration. 
The entry ``symmetry'' shows
which symmetry has been used to calculate the Finkelstein-Rubinstein
constraints. Here $(n,m)$ 
corresponds to the axial symmetry of the corresponding toroidal ansatz
(\ref{Utorus}). $C_2^0$ is generated by $\pi$ rotation
in space whereas $C_2^1$ is generated by
 rotation by $\pi$ in space followed
by rotation by $\pi$ in target space. As a word of caution, while axial
symmetry has been checked numerically, the $C_2$ symmetry is obtained by 
inspection from the figures in \cite{Hietarinta:2000ci} and
\cite{Battye:1998zn}. 
For low $Q$ the symmetries are apparent. However, for higher
Hopf charge, $Q>4$, the symmetries are difficult to guess, if indeed they
exist at all. Where no entry is given, the classical solution has no
obvious symmetry and the only constraint applicable is that of consistent
fermionicity. 

\begin{table}[!ht]
\begin{center}
\begin{tabular}{|l|l|l|l|l|l|l|l|}
\hline
& & & & & & &\\
$|Q|$ & $E_Q$ & shape & symmetry & FR & ground state & excited state (1) 
& excited state (2) \\
& & & & & & &\\
\hline
$1$     & 135.2 & unknot & $(1,1)$  & 1 &
$|\frac{1}{2},  - \frac{1}{2}, \frac{1}{2} \rangle$
&
$|\frac{3}{2},  - \frac{1}{2}, \frac{1}{2}\rangle$
&
$|\frac{3}{2},  - \frac{3}{2}, \frac{3}{2} \rangle$
\\
\hline
$2$     & 220.6 & unknot & $(2,1)$ & 1 &
$ | 0,0,0 \rangle$ 
&
$ | 1,0,0 \rangle$ 
&
$ | 2,- 2, 1 \rangle$ 
\\
\hline
$2^*$   & 249.6 & unknot & $(1,2)$ & 1 &
$ | 0,0,0 \rangle$ 
&
$ | 1,0,0 \rangle$ 
&
$ | 1,-1,1 \rangle$ 
\\
\hline
$3$ & 308.9 & unknot & $C_2^1$ & -1& 
$|\frac{1}{2}, \frac{1}{2}, \frac{1}{2} \rangle$
&
$|\frac{3}{2}, \frac{1}{2}, \frac{1}{2} \rangle$
& 
$|\frac{1}{2}, - \frac{1}{2}, \frac{3}{2} \rangle$
\\
\hline
$3^*$ & 311.3 & unknot & $(3,1)$ & 1 &
$|\frac{3}{2}, - \frac{3}{2}, \frac{1}{2} \rangle$
&
$|\frac{5}{2}, - \frac{3}{2}, \frac{1}{2} \rangle$
&
$|\frac{9}{2}, -\frac{9}{2}, \frac{3}{2} \rangle$
\\
\hline
$4$ & 385.5 & unknot & $(2,2)$ & 1 &
$ | 0,0,0 \rangle$ 
&
$ | 1,0,0 \rangle$ 
&
$ | 2,-2,1 \rangle$ 
\\
\hline
$4^*$ & 392.7 & unknot & $C_2^0$ & 1 &
$ | 0,0,0 \rangle$ 
&
$ | 1,0,0 \rangle$ 
&
$ | 0,0,1 \rangle$ 
\\
\hline
$4^*$ & 405.0 & unknot & $(4,1)$ & 1 &
$ | 0,0,0 \rangle$ 
&
$ | 1,0,0 \rangle$ 
&
$ | 4,-4,1 \rangle$ 
\\
\hline
$5$ & 459.8 & link & --- & --- &
$|\frac{1}{2}, \pm \frac{1}{2}, \frac{1}{2} \rangle$
&
$|\frac{3}{2}, \pm \frac{1}{2}, \frac{1}{2} \rangle$
&
$|\frac{1}{2}, \pm \frac{1}{2}, \frac{3}{2} \rangle$
\\
\hline
$5^*$ & 479.2 & unknot & $C_2^1$ & 1 & 
$|\frac{1}{2}, -\frac{1}{2}, \frac{1}{2} \rangle$
&
$|\frac{3}{2}, -\frac{1}{2}, \frac{1}{2} \rangle$
&
$|\frac{1}{2}, \frac{1}{2}, \frac{3}{2} \rangle$
\\
\hline
$6$ & 521.0 & link & --- & --- & 
$ | 0,0,0 \rangle$ 
&
$ | 1,0,0 \rangle$ 
&
$ | 0,0,1 \rangle$ 
\\
\hline
$6^*$  & 536.2 & link & --- & --- & 
$ | 0,0,0 \rangle$ 
&
$ | 1,0,0 \rangle$ 
&
$ | 0,0,1 \rangle$ 
\\
\hline
$7$ & 589.0 & knot & --- & --- & 
$|\frac{1}{2}, \pm \frac{1}{2}, \frac{1}{2} \rangle$
&
$|\frac{3}{2}, \pm \frac{1}{2}, \frac{1}{2} \rangle$
&
$|\frac{1}{2}, \pm \frac{1}{2}, \frac{3}{2} \rangle$
\\
\hline
\end{tabular}
\caption{Ground states and excited states for $Q=1,\dots,7$.
\label{table1}}
\end{center}
\end{table}

``FR'' gives the Finkelstein-Rubinstein constraints $(-1)^N$ where $N$
is calculated with equation (\ref{N}) for the generator of the
discrete symmetries.  
Note that axial symmetry
implies FR $= 1$. 
Then ground states are calculated as explained above. 
They are given in the form $|L L_3 K_3\rangle$. The quantum numbers $J_3$
and $I_3$ are suppressed. Recall that $J=L$ and $|I_3|=|K_3|$.
We have also included two excited states.
``Excited state (1)'', is obtained from the ground state by 
increasing $L$ by 1 and finding the lowest $K_3$
such that all constraints are satisfied. Similarly, ``excited state (2)''
is obtained by increasing $K_3$ by 1.
Note that changing the sign of ${\hat L}_3$ and ${\hat K}_3$ in the
constraints (\ref{FRconstraint}) given by a loop $L(\alpha, \beta)_\nvec$ can
be interpreted as constraints for the loop $L(-\alpha, -\beta)_\nvec$. 
Since the
fundamental group is ${\mathbb Z}_2$ the loop $L(\alpha, \beta)_\nvec$ is
contractible if and only if $L(-\alpha, -\beta)_\nvec$ is contractible. 
Therefore, whenever $|L, L_3, K_3 \rangle$ satisfies the constraints
imposed by a symmetry, 
so does $|L, -L_3, -K_3 \rangle$. In table \ref{table1}, we only display
states with $K_3 \ge 0$.

Since no constraints with FR $= -1$ occur for even Hopf charge $Q$ all
the ground states are given by $|0,0,0 \rangle$ and ``excited
states (1)'' are $|1,0,0 \rangle$. The influence of the
Finkelstein-Rubinstein constraints can 
only be seen for ``excited state (2)''. For odd $Q$ the
Finkelstein-Rubinstein constraints influence the ground states and all
the excited states.  

One might ask why the first and second excited states are expected to
have spin and isospin one unit higher than the ground state, respectively.
One reason is that this is consistent with the collective coordinate
quantization of Hopf solitons, to which we turn in the next section.


\section{Collective coordinate quantization}
\label{sec:collco}
\news

The simplest non-trivial quantitative 
application of our results is the collective
coordinate quantization, \cite{Su:2001zw}. In this case the wave function
is only non-vanishing on the space of minimal energy configurations in a
given sector, also called the moduli space. 
 The effective Lagrangian $\Leff$
in this approximation
is obtained by restricting the full Lagrangian
to fields which, at each fixed time, lie in the moduli space.\footnote{As 
has 
been discussed in the Skyrme model, \cite{Bander:gr, Braaten:1984qe}, this
approximation breaks
down if centrifugal effects are taken into account. This problem can be
avoided by introducing a (sufficiently large) mass term for the vector
$\nvec$ so that the fields decay fast enough at infinity.} From
$\Leff$ one can construct an effective Hamiltonian and canonically
quantize the system in the standard manner.
For Hopf charge $Q=1$ the reduced Hamiltonian is given in  
\cite{Su:2001zw} using ``$SU(2)$ notation''.

The Lagrangian  $ L$ (\ref{Lagrangian}) can be split up into
kinetic energy $T$ and potential energy $ V$,
namely $L = {T} - {V}$ where
\bea\label{Tdef}
{T} &=& \int_{\R^3}\, \frac{1}{2}|\partial_t \nvec|^2 
+ \frac{\lambda}{2}\sum_i|\partial_t \nvec \times \partial_i \nvec|^2, \\
\label{Vdef}
{V} &=& \int_{\R^3}\, \frac{1}{2}\sum_i|\partial_i \nvec|^2
+ \frac{\lambda}{4}\sum_{i,j}|\partial_i \nvec \times \partial_j \nvec|^2.
\eea
Now let $\ms\subset\Q_Q$ be the moduli space of charge $Q$ energy minimizers,
and $\nvec(t)$ be a trajectory in $\ms$. Since $\nvec(t)$ is a critical
point of $V$ for all $t$, $V$ must remain constant, $V[\nvec(t)]=M_0$ say,
interpreted as the classical mass of the Hopf soliton. It follows
that the effective Lagrangian is 
$\Leff= T|_\ms-M_0$, so the reduced dynamics 
is determined purely by the kinetic energy restricted to $\ms$. This has
a natural geometric interpretation:
being quadratic in first time derivatives, $T$ defines a positive
quadratic form
and hence a unique Riemannian metric $\gamma$ on $\ms$, and the classical
dynamics descending from $\Leff$ is nothing other than geodesic motion
in $(\ms,\gamma)$. Since the Faddeev-Hopf model
is not of Bogomol'nyi type, $\ms$ is just the orbit of any
energy minimizer under the symmetry group of the model, that is, 
all zero modes arise due to symmetry. The centre of mass motion decouples,
so we may, without loss of generality, assume that the centre of mass is
fixed at the origin, so that $\ms$ is the orbit of some minimizer
$\nvec_0$ under $G=SO(3)\times SO(2)$, acting as described in section 
\ref{Hopf}. So $(\ms,\gamma)$ is a homogeneous space, diffeomorphic to
$G/K$ where $K<G$ is the isotropy group of $\nvec_0$. It follows that
$\gamma$ is uniquely determined by its value on 
$T_{\nvec_0}\ms$.
 
Generically, as we have described, $K$ is discrete, so $\ms$ has dimension
4, and $\gamma$ is specified by 6 constants, which may be interpreted as
the components of the Hopf soliton's inertia tensor. However, we
shall concentrate on the case where $\nvec$ has axial symmetry. Then
\beq
\label{Kdef}
K=\{k(\alpha)=
([\diag(e^{i\alpha/2},e^{-i\alpha/2})],e^{in\alpha}):\alpha\in\R\}
\eeq
for some divisor $n$ of $Q$, where we have used the standard isomorphisms
$SO(3)\equiv PU(2)$ and $SO(2)\equiv U(1)$ to identify $SO(3)$ matrices with 
projective
equivalence classes of $U(2)$ matrices, and $SO(2)$ matrices with complex 
phases. Let $\theta_1,\theta_2,\theta_3$ be 
the usual basis of left invariant vector fields on $SO(3)$ and $\theta_4=
\cd_\xi$ on $SO(2)\equiv\{e^{i\xi}:\xi\in\R\}$. Let $\langle\cdots\rangle$
denote linear span.
Then the Lie algebra of
$G$, is $\g=\langle\theta_1,\ldots,\theta_4\rangle$, and the Lie algebra 
of $K$ is $\kk=\langle \theta_3+n\theta_4\rangle$. 
We may identify $T_{\nvec_0}\ms$
with the complementary space $\p=\langle\theta_1,\theta_2,
\theta_3\rangle$. Note $\g=\kk\oplus\p$ since $n\neq 0$. 
So $\gamma$ is equivalent to a positive symmetric
bilinear form $\bar{\gamma}:\p\oplus\p\ra\R$, and this must be invariant 
under the
adjoint action of $K$ on $\p$. Relative to the basis $\{\theta_1,\theta_2,
\theta_3\}$ this is
\beq
Ad_{k(\alpha)}=\left(\begin{array}{ccc}\cos\alpha&-\sin\alpha&0\\
\sin\alpha&\cos\alpha&0\\
0&0&1\end{array}\right).
\eeq
Let $\p^*$ denote the dual space to $\p$, so that $\bar{\gamma}\in
\p^*\odot\p^*$, where $\odot$ denotes the symmetric tensor product. The 
induced action of $K$ on $\p^*\odot\p^*$ may be 
decomposed into irreducible representations, whence one finds that
the dimension of the space of invariant symmetric bilinear forms on
$\p\oplus\p$ is \cite{jon}
\beq
\frac{1}{2\pi}\int_0^{2\pi}d\alpha\, \frac{1}{2}[(\tr Ad_{k(\alpha)})^2+
\tr(Ad_{k(\alpha)}^2)]=2.
\eeq
Hence there exist positive constants $a,b$ such that
\beq
\label{metric}
\bar\gamma=a(\sigma_1^2+\sigma_2^2)+b\sigma_3^2
\eeq
where $\{\sigma_i\}$ are the one forms dual to $\{\theta_i\}$. Thus the
metric $\gamma$ on $\ms$ is determined by just two constants.

The static solution $\nvec_0$, and hence its classical mass $M_0$ and 
moments of inertia $a$, $b$, all depend parametrically on the coupling 
$\lambda$. In fact, this dependence is quite simple, as we shall now show.
Let us temporarily denote all $\lambda$ dependence explicitly, so that
$T_\lambda$, $V_\lambda$ are the kinetic and potential energy functionals
at coupling $\lambda$, 
$\nvec_\lambda$ is the static solution, $M_0(\lambda)$ is its mass, and
$a(\lambda)$, $b(\lambda)$ its moments of inertia. A simple
rescaling of the integration variables in (\ref{Vdef}) shows that, for
any fixed map $\nvec:\R^3\ra S^2$,
\beq
V_\lambda[\nvec(\xvec)]\equiv \sqrt{\lambda}\, V_1[\nvec(\sqrt{\lambda}\,
\xvec)].
\eeq
Hence, given an extremal $\nvec_*$ of $V_1$ (here and henceforth, the
subscript 
$*$ will indicate that a quantity refers to the $\lambda=1$ model),
$\nvec_\lambda(\xvec)=\nvec_*(\lambda^{-\frac{1}{2}}\xvec)$ is an extremal
of $V_\lambda$, and furthermore its energy is
\beq\label{enscal}
M_0(\lambda)=V_\lambda[\nvec_\lambda]=\sqrt{\lambda}\, V_1[\nvec_*]=
\sqrt{\lambda}\, M_*.
\eeq
So the classical soliton masses scale as $\lambda^\frac{1}{2}$.
A similar argument works for the moments of inertia too. The coefficients
$a(\lambda)$, $b(\lambda)$ are, by definition, twice the kinetic energies
of the time-dependent fields, $\nvec_\lambda^{(i)}(\xvec,t)$ say, obtained
from $\nvec_\lambda$ by subjecting it to spatial rotation at unit
angular velocity about the $x_i$-axis with $i=1,3$ respectively. Let
$R_i(t)$ denote rotation through angle $t$ about the $x_i$-axis. Then
\beq
\nvec_\lambda^{(i)}(\xvec,t)=
\nvec_\lambda(R_i(t)\xvec)=
\nvec_*(\lambda^{-\frac{1}{2}}R_i(t)\xvec)=
\nvec_*(R_i(t)\lambda^{-\frac{1}{2}}\xvec)=
\nvec_*^{(i)}(\lambda^{-\frac{1}{2}}\xvec,t),
\eeq
by linearity of $R_i$. Rescaling the integration variables in (\ref{Tdef})
as before, one sees that $T_\lambda[\nvec_\lambda^{(i)}]=\lambda^\frac{3}{2}
T_1[\nvec_*^{(i)}]$, and so the moments of inertia scale as 
$\lambda^\frac{3}{2}$:
\beq\label{moiscal}
a(\lambda)=\lambda^\frac{3}{2}\, a_*,\qquad
b(\lambda)=\lambda^\frac{3}{2}\, b_*.
\eeq
Note that neither of these arguments appealed to axial symmetry, so the same
scaling behaviour applies to solitons with only discrete (for example,
trivial) symmetry groups, also. This includes the
scaling behaviour of the moment of inertia associated with isorotation
(where this no longer coincides with spatial rotation) because
\beq
\nvec_\lambda^{({\rm iso})}(\xvec,t)=R_3(t)\nvec_\lambda(\xvec) 
=R_3(t)\nvec_*(\lambda^{-\frac{1}{2}}\xvec)=
\nvec_*^{({\rm iso})}(\lambda^{-\frac{1}{2}}\xvec,t).
\eeq
From now on, we will no longer denote the $\lambda$ dependence explicitly,
but will
retain the $*$ subscript for quantities associated with the $\lambda=1$
model.

We wish to quantize
geodesic motion on $\ms$, which may be formulated as a Hamiltonian flow on
$T^*\ms$, within the framework of Finkelstein and Rubinstein. As it stands, 
there is a problem with this, however. As shown above, the fundamental group
of $\Q_Q$,
the topological sector containing $\ms$, is $\Z_2$, whereas $\pi_1(\ms)=
\Z_{2n}$, where $n$ is the divisor of $Q$ appearing in (\ref{Kdef}). A proof
of this is presented in the appendix. So $\pi_1(\ms)\neq\pi_1(\Q_Q)$ unless
$n=1$, and this type of axial symmetry occurs only for
$Q=1$ and the metastable
$Q=2$ state,
according to Hieterinta et al \cite{Hietarinta:2000ci}. 
Nevertheless, a fermionic collective coordinate
approximation is still possible, 
the
key point being that in all cases the $2\pi$ spatial rotation loop
has order 2 in $\pi_1(\ms)$. It is slightly unfortunate that this is true
independent of $Q$, that is, whether $Q$ is odd or even. For consistency
we must thus {\em choose} bosonic quantization for $Q$ even, it is not
imposed on us by the topology of $\ms$. This illustrates that 
collective coordinate quantization can be quite treacherous in the absence
of a good understanding of the topology of the full configuration space.
 
To construct the collective coordinate quantization
it is convenient to exploit the $n$-fold covering map
$\varrho:SO(3)\ra G/K$ which maps $g\in SO(3)$ to the coset $(g,1)K$, that
is, the left coset of $K$ containing $(g,1)\in G$.
Note that $\varrho$ commutes with the natural $SO(3)$ left actions on
$SO(3)$ and $\ms$. 
Geodesics in $(\ms,\gamma)$ are the images of geodesics in $(SO(3),
\varrho^*\gamma)$, where the lifted metric $\varrho^*\gamma$ is precisely
(\ref{metric}), but with $\sigma_i$ now interpreted as (global) left 
invariant one forms on $SO(3)$, rather than basis vectors in $\p^*$. The
Hamiltonian generating geodesic flow in $(SO(3),\varrho^*\gamma)$ is
\beq
H=\frac{1}{2a}(L_1^2+L_2^2)+\frac{1}{2b}L_3^2=\frac{1}{2a} |{\bf L}|^2+
\left(\frac{1}{2b}-\frac{1}{2a}\right)L_3^2
\eeq
where $L_i:T^*SO(3)\ra\R$ are the 
angular momenta corresponding to the
vector fields $\theta_i$ (the components of the moment map for the 
Hamiltonian action of $SO(3)$ on $T^*SO(3)$). Their Poisson bracket algebra
is well known: $\{L_1,L_2\}=L_3$ and cyclic permutations. We may now
quantize in the usual way, replacing classical angular
 momenta by $\hat{L}_i$, self-adjoint
linear operators on $L^2(SO(3))$ and Poisson brackets by
commutators. Note that $\{\hat{H},\hat{\bf L}^2,\hat{L}_3\}$ is a compatible
set of observables.
In this set-up, we are 
thinking of the wavefunction as defined on
the covering space, $\psi:SO(3)\ra\C$; it is important to note that for
$Q$ odd (even)
 only those functions which
are double-valued (single-valued) under the projection $\varrho$ make 
physical 
sense. The 
deck transformation group for $\varrho$ is generated by 
$\exp(2\pi\theta_3/n)$, so we find that the eigenvalues of $\hat{L}_3$ must
be integer multiples of $n/2$. This conclusion may be reached another way.
Note that $\theta_3+n\theta_4\in\kk$ vanishes on $\ms$, so the
corresponding classical momenta are linearly dependent: $L_3+nK_3=0$.
Hence the quantum operators must satisfy $(\hat{L}_3+n\hat{K}_3)\psi=0$ on
any physical state, and the conclusion follows because
$\hat{K}_3$ has half-integer spectrum.
Of course, this is nothing other than the
FR constraint for axial symmetry (\ref{Caxial}). 
We may use the linear dependence of the third components of 
spin and isospin to rewrite $\hat{H}$ in terms of $\hat{K}_3$, or
both $\hat{L}_3$ and $\hat{K}_3$ if we wish. A convenient way to
write the quantum hamiltonian is
\begin{equation}
{\hat H} = M_0 + \frac{1}{2a} {\hat {\bf L}}^2 + 
\left( \frac{1}{2b} - \frac{1}{2a} \right) {\hat L}_3^2.
\end{equation} 
It is now trivial to express the quantum energy spectrum in terms of
the quantum numbers $L^2$ and $K_3$:
\beq
\label{ELK}
E 
=\sqrt{\lambda}\,
M_*+\frac{\hbar^2}{\lambda^\frac{3}{2}}\left[\frac{L(L+1)}{2a_*}+
\left(\frac{1}{2b_*}-\frac{1}{2a_*}\right)n^2K_3^2\right]
\eeq
where we have used the constraint $L_3 = -n K_3$ to eliminate $L_3$,
and the scaling behaviour obtained
in (\ref{enscal}),(\ref{moiscal}) to render all $\lambda$ dependence explicit.
Recall that $*$-subscript quantities refer to the $\lambda=1$ soliton.
As discussed in the previous section, the body-fixed and space-fixed
angular momenta satisfy  ${\hat {\bf J}}^2 = {\hat {\bf L}}^2$ and
${\hat I}_3^2 = {\hat K}_3^2$. Therefore, we can also express the
energy in terms of the space-fixed angular momentum quantum numbers,
which are the quantities measured in a physical experiment, by
replacing $L(L+1)$ by $J(J+1)$ and $K_3^2$ by $I_3^2$ in formula
(\ref{ELK}).

We would like to order these states by increasing energy. Clearly, this
order depends on $n$ and
the relative size of the constants $a_*$ and $b_*$. As discussed above,
to determine these constants, one must compute the kinetic energy
of time dependent fields $\nvec(t)=(\exp(t\theta_1),1)\cdot\nvec_0$
and $\nvec(t)=(\exp(t\theta_3),1)\cdot\nvec_0$ respectively,
where $\cdot$ denotes the action of $G$ on $\ms$. This is 
computationally very expensive if one uses for $\nvec_0$ the genuine
axially symmetric energy minimizers found in \cite{Hietarinta:2000ci},
since even to construct $\nvec_0$ requires one to solve nonlinear PDEs. 
Instead, we shall again exploit the Hopf fibration and assume that
$\nvec_0$ is well approximated by the image under the Hopf map 
$\rho$ of a Skyrme 
configuration $U:\R^3\ra SU(2)$ within the rational map ansatz of Houghton,
Manton and Sutcliffe \cite{Houghton:1998kg}. This idea
was introduced in \cite{Battye:1998zn}.\footnote{Su has also discussed the 
rational map ansatz for Hopf solitons, using a different notation, 
\cite{Su:2001yg}.} The rational map ansatz may be
described as follows. Using $\exp:\mathfrak{su}(2)\ra SU(2)$, one may
identify $SU(2)$ with the closed ball of radius $\pi$ in
$\mathfrak{su}(2)\equiv\R^3$. The entire boundary of this ball gets mapped
to $-\I_2$. Partition physical space $\R^3$ into concentric 2-spheres of
radius $r\in [0,\infty)$. Choose a fixed holomorphic map $
{\bf R}:S^2\ra S^2\subset\R^3$ of
degree $Q$ and a smooth decreasing surjection 
$f:[0,\infty)\ra(0,\pi]$ (the profile function). Then the corresponding
degree $Q$ Skyrme configuration  
is 
\beq
U(r,x_1,x_2)=\exp\left(f(r){\bf R}(x_1,x_2)\right)
\eeq
where
$x_1,x_2$ is any coordinate system on $S^2$. With respect to stereographic
coordinates $z$, $R$ on its domain and codomain, ${\bf R}$ is 
the eponymous rational map $R(z)$. We may then write $U(r,z)$ more
explicitly as
\begin{equation}
U(r,z) = \frac{1}{1+|R|^2}\left(
\begin{array}{cc}
e^{-if}+|R|^2e^{if} &
2i {\bar R} \sin f \\
2i R \sin f & 
e^{if}+|R|^2e^{-if}
\end{array}
\right).
\end{equation}
The corresponding Faddeev-Hopf configuration $\pi\circ U$ can easily
be calculated with equations (\ref{Hopfmap}) and (\ref{SU(2)id}), 
\beq
W(r,z)=\frac{|R(z)|^2e^{if(r)}+e^{-if(r)}}{2iR(z)\sin f(r)},
\eeq
where again we choose stereographic coordinates on $S^2$. The idea
is to approximate the true energy minimizer $\nvec_0$ by a configuration
of this form and minimize over all possible $R$ and $f$. In fact, to
obtain axial symmetry, we must assume $R(z)=z^Q$ (note this assumes the
divisor $n$ of $Q$ is simply $n=Q$, so our results apply only to $Q=1,2$
and the metastable $Q=3^*,$ $4^*$ solitons). We then minimize the potential
energy $V$ over all possible 
profile functions $f$. This yields a nonlinear second order ODE
for $f(r)$ which is easily solved numerically. We may, without loss of 
generality, set $\lambda$ to unity.

\begin{table}
\begin{center}
\begin{tabular}{|c|c|c|c|c|c|c|}
\hline
&&&&&&\\
$Q$& $M_*$& $M_*^H$& $M_*^G$& $M_*^B$&   $a_*$ & $b_*$ \\ 
&&&&&&\\
\hline
&&&&&&\\
$1$ & 275.0 & 270.4 & 278.6 & 252.5 & 418.8 & 369.7 \\
&&&&&&\\
$2$& 462.9 & 441.2 & 446.9  & 418.0 &  1265.0 & 1309.4 \\
&&&&&&\\
$3^*$& 665.5 & 622.6 &---& 590.5 & 3272.7 & 3556.1 \\ 
&&&&&&\\
\hline
\end{tabular}
\caption{Classical energy $M_*$ and moments of inertia $a_*$, $b_*$ of 
various axially symmetric solitons, at $\lambda=1$, within the
rational map ansatz. For comparison, we also quote the classical
energies of the corresponding numerical solutions found in the literature
($M_*^H$: Hietarinta and Salo
\cite{Hietarinta:2000ci}, $M_*^G$: Gladikowski and Hellmund
\cite{Gladikowski:1997mb}, $M_*^B$: 
Battye and Sutcliffe \cite{Battye:1998pe}). Note that $M_*^H$ and $M_*^G$
have been inferred using the scaling rule (\ref{enscal}).}\label{table3}
\end{center}
\end{table}

Having constructed our approximate energy minimizer, 
$W(r,z)$, we must compute
the kinetic energy at $t=0$ of
\beq
W(t,r,z)=\frac{|R(\tilde{z}(t,z))|^2e^{if(r)}+
e^{-if(r)}}{2iR(\tilde{z}(t,z))\sin f(r)}
\eeq
where 
\beq
\tilde{z}(t,z)=\frac{z\cos t/2+i\sin t/2}{iz\sin t/2+\cos t/2},\quad
\mbox{and}\quad
\tilde{z}(t,z)=ze^{it},
\eeq
yielding $a_*/2$ and $b_*/2$ respectively. The calculations are elementary,
but lengthy, and all reduce to radial integrals of expressions involving
$f(r)$ and $f'(r)$. The results for $Q=1,$ $2$ and the metastable
$Q=3^*$ are summarized in table \ref{table3}. These data, along with
formula (\ref{ELK}) give the complete quantum energy spectrum for these
solitons, at arbitrary coupling.

\begin{table}
\begin{center}
\begin{small}
\begin{tabular}{|c|c|c|c|c|c|c|}
\hline
& & & & & &\\
$Q$ & groundstate & $E_0$ & excited state (1) & $E_1$ &
excited state (2) & $E_2$ \\
& & & & & &\\
\hline
& & & & & &\\
$1$ & $|\frac{1}{2}, -\frac{1}{2}, \frac{1}{2} \rangle$ &
$6.63$ TeV &
$| \frac{3}{2}, -\frac{1}{2}, \frac{1}{2} \rangle$ &
$9.67$ TeV &
$| \frac{3}{2}, -\frac{3}{2}, \frac{3}{2} \rangle$ &
$9.93$ TeV \\
& & & & & &\\
\hline
& & & & & &\\
$2$ & $|0, 0, 0 \rangle$ &
$9.82$ TeV &
$| 1, 0, 0 \rangle$ &
$10.49$ TeV &
$| 2, -2, 1 \rangle$ &
$11.79$ TeV \\
& & & & & &\\
\hline
& & & & & &\\
$3^*$ & $|\frac{3}{2}, -\frac{3}{2}, \frac{1}{2} \rangle$ &
$14.58$ TeV &
$| \frac{5}{2}, -\frac{3}{2}, \frac{1}{2} \rangle$ &
$15.23$ TeV &
$| \frac{9}{2}, -\frac{9}{2}, \frac{3}{2} \rangle$ &
$17.12$ TeV \\
& & & & & &\\
\hline
\end{tabular}
\caption{Groundstates and first excited states, and their energies, 
of super heavy smoke ring solitons in
  the collective coordinate approximation, using the rational map
  ansatz.\label{table2}}
\end{small}
\end{center}
\end{table}

To illustrate our approach we shall interpret the Hopf solitons as super 
heavy fermion states in the strongly coupled pure Higgs sector of the 
standard model, as advocated by Gipson and Tze \cite{giptze}. To make
contact with their work, we must take the unit of energy to be $e_0=300$ GeV,
$\hbar=1$, and the coupling constant to be
$\lambda=\ln(m_H/e_0)/24\pi^2$, where $m_H$ is the Higgs mass. In
this model, the Higgs sector is strongly coupled, so the Higgs mass
assumes the rather large value $m_H\approx 1$ TeV, so
that $\lambda\approx 0.005$. The unit of length is the Compton wavelength
of a particle of rest energy $e_0$, namely $d_0=\hbar c/e_0
\approx 0.66~10^{-3}$ fm.
Then the $Q=1$ ground state represents what Gipson and Tze call a 
``smoke ring soliton'' of energy $6.63$ TeV which is compatible with
the lower bound of $5.5$ TeV given in \cite{giptze}. 
A sensible measure for the
size of the Hopf soliton is the value of the radius in the
rational map ansatz at which the profile function takes the value 
$\pi/2$.  	
We find that our Hopf soliton has a radius of $0.08~10^{-3}$ fm which is
comparable with the lower bound of $0.2~10^{-3}$ fm in \cite{giptze}
where the radius is defined in a slightly different way.
We display the groundstates and the first two
excited states in the collective coordinate approximation in table
\ref{table2}. The energies of the states are dominated by the
classical contribution. As anticipated in table \ref{table1}, 
the groundstate has the lowest
energy followed by excited state (1) and excited state (2). The energy
of the states increases with the Hopf charge $Q$.  
The size of the Hopf solitons also increases with the charge; 
$0.08~10^{-3}$ fm for $Q=1$, $0.09~10^{-3}$ fm  for $Q=2$ and
$0.13~10^{-3}$ fm for $Q=3$.

Clearly, the relative size of the quantum excitation energy of
an excited state to the ground state energy depends on the coupling 
$\lambda$. If $\lambda$ is small, as in the application above, the
quantum corrections become significant.
In an application where the solitons are taken to model
real physical structures, whose energies and sizes are known experimentally
(rather than hypothetical exotic matter states as in the current case), one
would tune the energy and length scales independently so as to fit some
reference data as well as possible. This amounts to tuning both $\lambda$
and the value of $\hbar$, which is why we retained explicit $\hbar$ 
dependence in equation (\ref{ELK}). 
In the case of the Skyrme system as a model
of nucleons, 
for example, one finds that $\hbar\approx 46.8$ in natural units 
\cite{Leese:1994hb}. Even if $\lambda$ is large, therefore, quantum 
corrections may still be significant, provided
$\hbar/\lambda$ remains large.
So the relative importance of quantum corrections depends strongly on
the physical interpretation of the model under consideration.

\section{Conclusion}
\label{sec:conc}
\news

We have described how to quantize Hopf solitons using the
Finkelstein-Rubinstein construction and thereby demonstrated that 
Hopf solitons can be quantized as fermions when their Hopf charge
$Q$ is odd. An important ingredient of the proof is the fact
that the Hopf map 
$S^3 \to S^2$ induces a Serre fibration $(S^3)^M\ra (S^2)^M_*$.
Using this fibration we could show that the fundamental group of Skyrmions
is isomorphic to the fundamental group of Hopf solitons, when physical
space has finite fundamental group, and this
isomorphism is induced by the Hopf map. This enabled us to use results
which have been derived for the Skyrme model. 

In a semiclassical quantization we expect that classical symmetries are
not broken by quantum effects. Then the symmetries of the classical
configurations induce non-trivial constraints on the wave
function. We calculated possible ground states of Hopf solitons for $Q=1,
\dots, 7$
from the minimal energy configurations given in \cite{Hietarinta:2000ci}.
Since Hopf solitons do not have many symmetries, the constraints on the
wave functions are quite weak. Often, only the degeneracy of a state
changes, rather than the state being excluded completely. 
Excited states have been included to better
illustrate the influence of the Finkelstein-Rubinstein constraints.

In order to get quantitative predictions of the quantum energy spectrum
of Hopf solitons, we resorted to a collective coordinate 
approximation. In general, naive collective coordinate
quantization can give spurious results if the
topology of the moduli space is incompatible with that of 
the full configuration space. 
We concentrated on the case
 where the moduli space consists of axially symmetric
configurations, which provides a good example of this difficulty. As
discussed in the previous section, such a moduli space allows
for fermionic quantization for both odd and even Hopf charge. 
In order
to describe the physics correctly, we have to impose bosonic
quantization for even $Q$ and fermionic quantization for odd $Q$. In other
words, we must impose some of
the Finkelstein-Rubinstein constraints arising from
the topology of the full configuration space ``by hand''
on the wave function on the
moduli space. They do not arise from the topology of the moduli space
itself. 

The Faddeev-Hopf model contains a single coupling constant $\lambda$. 
By simple
rescaling arguments, we derived the scaling behaviour of the classical
energy and moments of inertia of a soliton as $\lambda$ varies. 
This allowed us to find a formula for the quantum energy spectrum of
axially symmetric solitons, within
the collective coordinate approximation, with
all $\lambda$ dependence explicit. The numerical constants $M_*$, 
$a_*$ and $b_*$ in this formula were approximated, for three such axially
symmetric solitons, by constructing approximate energy minimizers within
the rational map ansatz. Our aim in this paper was to illustrate the general
approach of fermionic soliton quantization within the Faddeev-Hopf model. 
This can now be applied to a
variety of physical models that admit Hopf solitons.

\section*{Acknowledgements} The authors wish to thank D Auckly and 
P M Sutcliffe for
fruitful discussions. S K acknowledges an EPSRC Research fellowship
GR/S29478/01.

\appendix
\section*{Appendix: The fundamental group of the moduli space}

We wish to compute the fundamental group of
$\ms$, the orbit of a configuration $\nvec:\R^3\ra S^2$
under $G=SO(3)\times SO(2)$, when $\nvec$ is invariant under the axial
symmetry group $K=\{(R_3(\alpha),e^{in\alpha})\, :\, \alpha\in\R\}<G$, 
where $R_3(\alpha)$
denotes rotation through $\alpha$ about the $x_3$ axis. Since $\ms\cong
G/K$ and $p:G\ra G/K$ is a fibration, we have the associated homotopy exact
sequence
$$
\begin{array}{rcccccc}
K&\stackrel{\iota}{\hookrightarrow}&G&\stackrel{p}{\ra}&G/K&&\\
\Rightarrow\quad
\pi_1(K)&\stackrel{\iota_*}{\ra}&\pi_1(G)&\stackrel{p_*}{\ra}&\pi_1(\ms)&
\ra&\pi_0(K)\\
\Z&\stackrel{\iota_*}{\ra}&\Z_2\oplus\Z&\stackrel{p_*}{\ra}&\pi_1(\ms)&
\ra&0.
\end{array}
$$
Hence $p_*$ surjects, so $\pi_1(\ms)\equiv\pi_1(G)/\ker p_*$ by the 
Isomorphism Theorem. But $\ker p_*$ is, by exactness, the image of $\pi_1(K)$
under inclusion, clearly the infinite
cyclic group generated by $1\oplus n\in
\pi_1(G)$. This group has precisely $2n$ cosets in $\pi_1(G)$, labelled by
the elements 
$$
0\oplus 0,\, 0\oplus 1,\, \ldots,\, 0\oplus(2n-1),
$$
 for example. 
Let us denote the coset $g+\ker p_*$ by $[g]$. It
follows immediately that the quotient group $\pi_1(G)/\ker p_*$ is cyclic
of order $2n$, generated by $[0\oplus 1]$. 
Note also that the $2\pi$ spatial rotation loop lies in
$1\oplus 0\in\pi_1(G)$, which projects to $[0\oplus n]=n[0\oplus 1]$ in
$\pi_1(G)/\ker p_*$, since  $1\oplus 0=0\oplus n-1\oplus n$. Hence the
$2\pi$ spatial rotation loop in $\ms$ is noncontractible of order 2, 
independent of $n$ (and $Q$).

\end{document}